\documentstyle[multicol,prl,aps]{revtex}
\input epsf
\def\DSAS{\Delta_{\text{SAS}}}

\def\dnA{\downarrow}
\def\nt{n_{\text{t}}}

\def\etal{{\it et~al.\ }}
\begin{document}
\titlepage
%%%%%%%%%%%%%
\title{
Interlayer Coherence in the $\nu=1$ and $\nu=2$
Bilayer Quantum Hall States
}
%%%%%%%%%%%%%
\author{
A. Sawada$^{\rm 1}$, Z.F. Ezawa$^{\rm 1}$, H. Ohno$^{\rm 2}$, 
Y. Horikoshi$^{\rm 3}$, A. Urayama$^{\rm 1}$\\
Y. Ohno$^{\rm 2}$, S. Kishimoto$^{\rm 2}$, F. Matsukura$^{\rm 2}$, 
N. Kumada$^{\rm 1}$
}

\address{$^1$Department of Physics, Tohoku University, Sendai 980-8578, Japan}

\address{$^2$Research Institute of Electrical Communication,
Tohoku University, Sendai 980-8577, Japan}

\address{$^3$School of Science and Engineering, 
Waseda University, Tokyo 169-8555, Japan}

\vspace{5mm}

%%%%%%%%%%%%%
\maketitle
%%%%%%%%%%%%%
\begin{abstract}
We have measured the Hall-plateau width and the activation energy of the bilayer quantum Hall (BLQH) states at the Landau-level filling factor ${\nu}=1$ and 2 by tilting the sample and simultaneously changing the electron density in each quantum well. The phase transition between the commensurate and incommensurate states are confirmed at $\nu =1$ and discovered at $\nu =2$. In particular, three  different $\nu =2$ BLQH states are identified; the compound state, the coherent commensurate state, and the coherent incommensurate state.\\ \\
\end{abstract}
%%%%%%%%%%%%%

\vspace*{-7mm}
\begin{multicols}{2}\narrowtext
\setcounter{page}{1}
A spontaneous development of interlayer quantum 
coherence \cite{EIcoher,WenZee} is one of 
the most interesting phenomena in bilayer quantum Hall (BLQH) systems.  
One can experimentally prove the existence of such 
an interlayer quantum coherence 
by manipulating the macroscopic quantum conjugate observables;
the phase difference and the interlayer electron number difference.
The interlayer phase difference, if exists, 
can be controlled by applying a parallel magnetic field 
between the two layers, 
which can be achieved by tilting the bilayer system in a magnetic field. 
Murphy {\it et al.} \cite{Sheena} have found an activation-energy
anomaly together with a phase transition 
in the $\nu=1$ BLQH state by increasing the parallel magnetic field.
It was suggested to be a signal of the interlayer quantum coherence
\cite{EIplasmon,YangMoon}. 
The interlayer number difference can also be controlled experimentally 
by applying gate bias voltages to the two layers.
When the interlayer coherence exists, 
the BLQH state persists even if the electron density is arbitrarily 
unbalanced between two quantum wells.
Sawada {\it et al.} \cite{SawaPRLa} have found precisely 
this behavior in certain BLQH states; 
furthermore they have found that the activation energy increases
as the density difference becomes larger.
This behavior is presumably due to a capasitive charging energy stored 
in Skyrmions excited across the two layers 
in the coherent state \cite{EzaICBa}.
These two experiments \cite{Sheena,SawaPRLa} indicate strongly 
the spontaneous development of the interlayer coherence in
the $\nu=1$ BLQH state.

An intriguing problem is whether an interlayer coherence
develops also in the $\nu=2$ BLQH systems.
Sawada \etal have found \cite{SawaPRLa} a phase transition 
at $\nu=2$ by changing the electron density continuously.
The phase transition occurs seemingly between the $\nu=1+1$ compound state 
and the $\nu=2$ ``coherent state''.  
Phase transitions have also been observed at $\nu=2$ 
in optical experiments by Pellegrini {\it et al.}:
first they used samples with different densities\cite{PellegriniA};  
second they tilted samples in the magnetic field \cite{PellegriniB}. 
In each of them, they have found two distinctive phases 
with respect to spin-excitation modes.
They have concluded a phase transition between spin polarized and unpolarized
states at $\nu=2$. 
It is important to relate the phase transitions
found in the magnetotransport \cite{SawaPRLa} 
and optical \cite{PellegriniA,PellegriniB} experiments, if any.

In this paper, we report the results of experiments on the $\nu =1$ 
and 2 BLQH states, 
where we have measured the Hall-plateau width and the 
activation energy by changing the density in each quantum well and 
simultaneously tilting the sample in a magnetic field.  
In this way we control both the density difference and 
the conjugate phase difference simultaneously in a BLQH state.  

The sample was grown by molecular beam epitaxy on a (100)-oriented GaAs 
substrate, and consists of two modulation doped GaAs quantum wells of width 
$W=200$\,\AA, separated by an Al$_{0.3}$\-Ga$_{0.7}$\-As barrier of thickness 
$d_{\text{B}}=31$\,\AA.  
The total electron density $n_{\text{t}}$ of this sample was 
$2.3\times 10^{11}\,$cm$^{-2}$ at zero gate voltage, the mobility was 
$3.0\times 10^5\,$cm$^2$/Vs at temperature $T=30$\,mK, and  
the tunneling-energy gap $\DSAS\ $ was 6.8 \,K.  
The Schottky gate electrodes were fabricated on 
both front and back surfaces of the sample so 
that the front-layer ($n_{\rm f}$) 
and the back-layer electron density 
($n_{\rm b}$) can be independently controlled by 
adjusting the front ($V_{\rm fg}$) and the back gate voltage ($V_{\rm bg}$).

Measurements were performed with the sample mounted in a mixing chamber 
of a dilution refrigerator.  
The magnetic field with maximum 13.5\,T was applied to the sample.  
Standard low-frequency ac lock-in techniques were used 
with currents less than 100\,nA to avoid heating effects.  
The sample mounted on a goniometer with 
the superconducting stepper motor \cite{GONO} can be rotated 
into any direction in the magnetic field.

The Hall-plateau width has previously been shown 
to be a good indicator of the stability of the QH state,
and the correlation with the activation energy has been 
established \cite{SawaPRLa}.
Its dependence on the tilted angle $\Theta$ 
and the normalized density difference 
$\sigma=(n_f-n_b)/n_t$
gives an overview in categorizing different types of BLQH states.

%%%%%%%%%%%%%%%%%%%%%%%%%%%%%
\begin{figure}[!bth]
\epsfxsize=80mm
\epsfbox{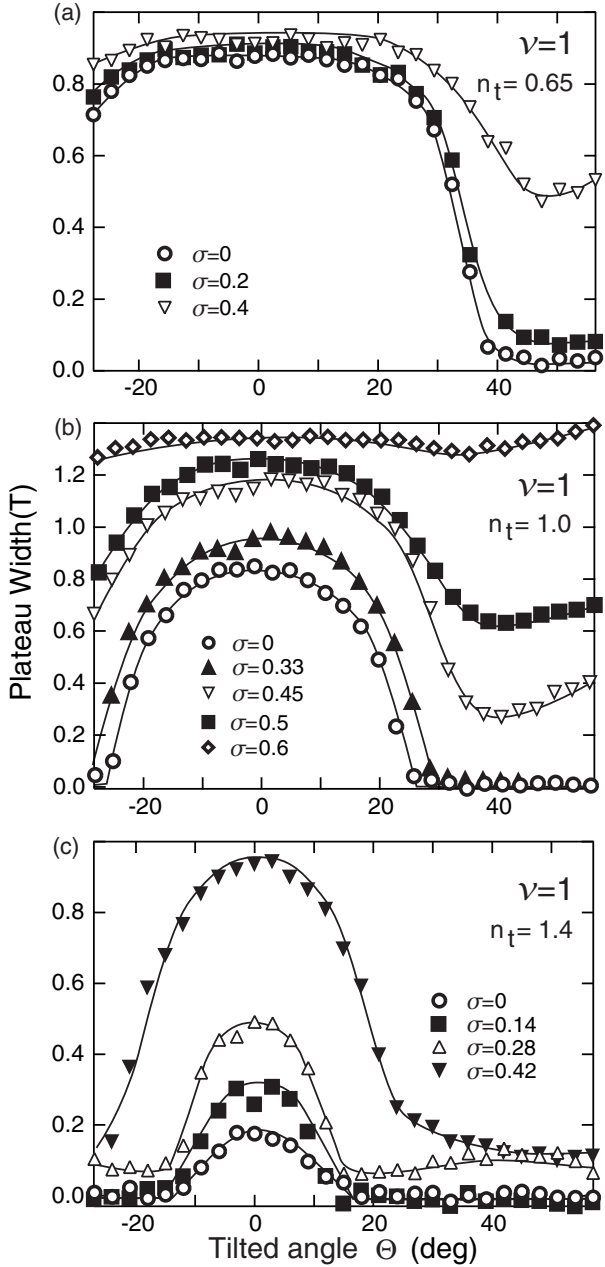}
\caption{\small
The Hall-plateau width of the ${\nu}=1$ state at 50\,mK as a function of the 
tilted angle $\Theta $ at various $\nt$ and $\sigma $.}
\label{Plateau1}
\end{figure}
%%%%%%%%%%%%%%%%%%%%%%%%%%%%%

In Fig.~\ref{Plateau1} we show the plateau width of the $\nu =1$ BLQH 
state as a function of $\Theta$ at various $n_{\rm t}$ and 
$\sigma$.  
The plateau width is defined \cite{SSC} with 
respect to the perpendicular field $B_{\perp }$.  
All data of the plateau width exhibit a similar behavior.  

\begin{figure}[!bth]
\epsfxsize=80mm
\epsfbox{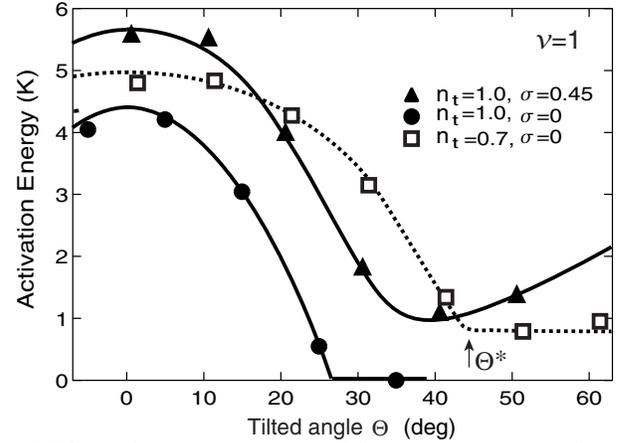}
\caption{\small
Activation energy of the $\nu=1$ state
as a function of $\Theta$ at various $n_{\rm t}$ and $\sigma$.  The total 
density $n_{\rm t}$ is in unit of $10^{11}$\,cm$^{-2}$.  
}
\label{Activa1}
\end{figure}

We give the activation energy as a function of $\Theta $ in 
Fig.\ref{Activa1}.  
As typical examples we show the data with 
$\nt=1.0$ and 0.7 in unit of $10^{11}$\,cm$^{-2}$.  
Two curves are at the 
balanced point ($\sigma =0$) and one at 
the unbalanced point ($\sigma =0.45$).  
The activation energy $\Delta $ is derived from 
the temperature dependence of the magnetoresistance; 
$R_{\rm xx}=R_0 \exp (-\Delta /2T)$.  
(This definition is 
different by factor 2 from the previous one \cite{SawaPRLa}.)  

The activation energy has a peak at $\Theta =0$, 
and drops rapidly to a certain tilted angle $\Theta ^{*}$,  
and then it becomes flat ($\sigma =0$) or increases ($\sigma \not=0$).  
This behavior is the anomaly revealed first by Murphy \etal \cite{Sheena} at 
the balanced point ($\sigma =0$).  
The critical angle $\Theta ^{*}$ clearly indicates a phase transition.  
Yang \etal \cite{YangMoon} have argued that 
it is the commensurate state for $\Theta <\Theta ^{*}$ and 
the incommensurate state for $\Theta >\Theta ^{*}$, 
about which we explain later based on eq.(\ref{EffecHamil}).  
We also identify $\Theta ^{*}$ with the 
commensurate-incommensurate (CIC) phase transition point.  
Our new finding is that the CIC transition occurs also 
in unbalanced configurations ($\sigma \not=0$).  
As we show later in eq.(\ref{PhaseAngle}), 
the phase difference $\theta $ is related to 
$\Theta $ in the interlayer coherent phase.  
Thus, each BLQH state turns out to possess definite values of 
$\sigma $ and $\theta $ in Fig.\ref{Plateau1}.  
We conclude that this is an 
evidence of the development of the interlayer coherence at $\nu =1$.

%%%%%%%%%%%%%%%%%%%%%%%%%%%%%
%%%%%%%%%%%%%%%%%%%%%%%%%%%%%

\begin{figure}[tbh]
\epsfxsize=80mm
\epsfbox{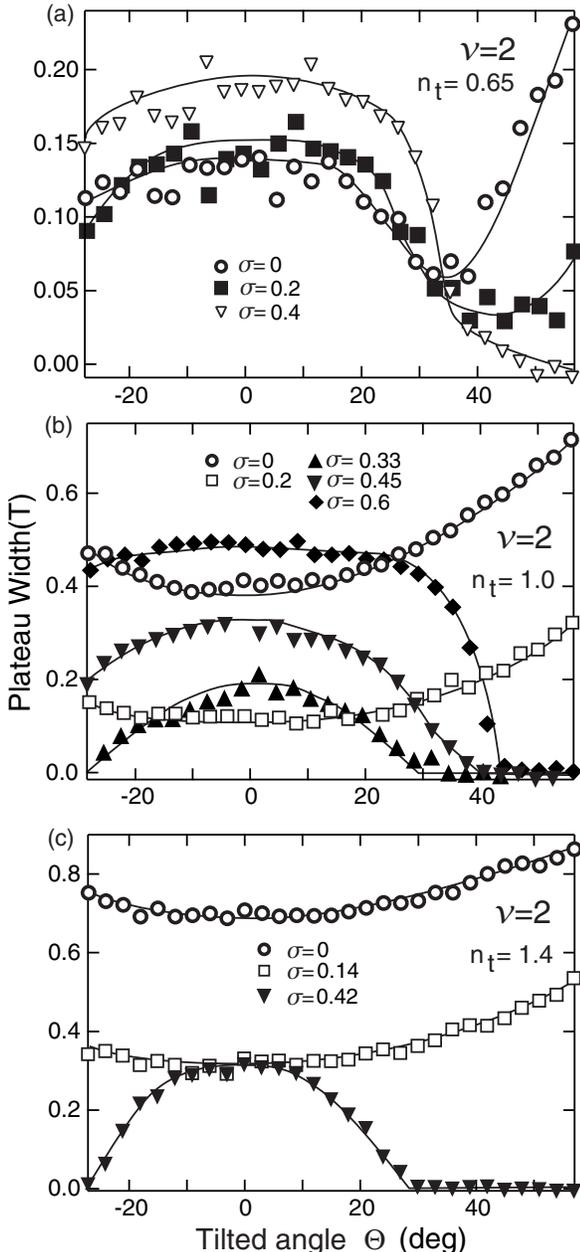}
\caption{\small
The Hall-plateau width of the ${\nu}=2$ state at 50\,mK
as a function of $\Theta$ at various $\nt$ and $\sigma $.
}
\label{Plateau2}
\end{figure}

We next show the plateau width of the $\nu =2$ BLQH state in 
Fig.\ref{Plateau2}.
There are two distinguished behaviors, 
as is consistent with the previous data \cite{SawaPRLa}:
(A) The overall behavior at a low density [Fig.\ref{Plateau2}(a)] bears a 
close resemblance to that in the $\nu=1$ state (Fig.\ref{Plateau1}).  
It indicates that the interlayer coherence has developed also at $\nu =2$ 
together with the CIC transition.
(B) At higher densities [Fig.\ref{Plateau2}(b) and (c)], 
we observe two distinguished types of states:  
(B1) The plateau width near the balanced point ($\sigma \simeq 0$) 
increases monotonously as the tilted angle increases; 
(B2) The plateau width at large off-balanced points shows 
a behavior characteristic to the coherent state.  

\begin{figure}[!bth]
\epsfxsize=80mm
\epsfbox{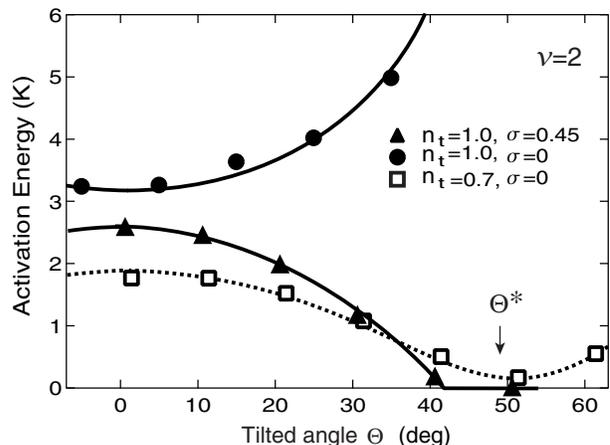}
\caption{\small
Activation energy of the $\nu=2$ QH state
as a function of $\Theta$ at various $n_{\rm t}$ and $\sigma$.
}\label{Activa2}
\end{figure}

We give the activation energy as a function of $\Theta $ in 
Fig.\ref{Activa2}, where $\nt=1.0$ and 0.7 in unit of $10^{11}$ cm$^{-2}$.  
(A) At low density ($\nt=0.7$) it shows an anomalous behavior 
in the activation energy 
as in the $\nu =1$ coherent BLQH state.  
However, the activation energy begins to increase beyond 
$\Theta ^{*}$, 
whose origin will be the Zeeman energy of spin excitations 
as we discuss later  
(see Table\ 1).
(B1) At higher density ($\nt=1.0$) 
it increases monotonically at the balanced point ($\sigma =0$).  
This is an expected behavior in the compound state 
which is stable only around the balanced point \cite{SawaPRLa}.  
The increase is due to the Zeeman energy of spin excitations.  
Note that no tunneling energy contributes to the compound state.
(B2) At the off-balanced point ($\sigma =0.45$) 
its behavior is that of a typical 
coherent state established in the $\nu =1$ BLQH state.  

%%%%%%%%%%%%%%%%%%%%%%%%%%%%%
%%%%%%%%%%%%%%%%%%%%%%%%%%%%%

We proceed to discuss physics behind the interlayer coherence of the 
bilayer QH states.  
The interlayer coherence is described by the Hamiltonian density 
\cite{YangMoon,EzaIQC},
%%%%%%%%%%%%%%%%%%%%%%%%%%%
\begin{eqnarray}
{\cal H}= 
{\frac{\rho _{\rm s}}{2}} [(\partial _{\rm x}\theta )^{2}+(\partial _{\rm x}\sigma )^{2}] 
+ {\frac{e^{2}\nt^{2}}{8C}} \sigma^{2}  \nonumber \\
- {\frac{\DSAS \nt}{4}} 
\sqrt{1-\sigma^2}
\cos(\theta -Qx),  
\label{EffecHamil}
\end{eqnarray}
%%%%%%%%%%%%%%%%%%%%%%%%%%%
where $Q=2\pi dB_{\parallel }/\phi _{\rm 0}$ with 
the Dirac flux unit $\phi _{\rm 0}\equiv h/e$.
We have taken the BLQH 
system parallel to the $xy$ plane and applied the parallel magnetic field to 
the $y$ direction.  
The first term describes the Coulomb exchange energy with 
the pseudospin stiffness 
$\rho _{\rm s}\simeq \nu e^{2}/(16\sqrt {2\pi }\varepsilon \ell _{\rm B})$; 
the second term the capacitive 
charging energy with the capacitance $C$, and the last term the tunneling 
energy.  
($\varepsilon $ is the dielectric constant.)  
The tilted angle $\Theta$ is given by 
$\tan\Theta =B_{\parallel }/B_{\perp }$.  
The phase difference $\theta$ induces screening currents 
$J^{\text{f}}_{\rm x} = -J^{\text{b}}_{\rm x} 
= (2\pi e\rho _{\rm s}/h)\partial _{\rm x}\theta$
on the two layers into the opposite directions \cite{EIplasmon}.

%%%%%%%%%%%%%%%%%%%%%%%%%%%%
%%%%%%%%%%%%%%%%%%%%%%%%%%%%%

On one hand, 
in the commensurate phase ($\Theta <\Theta ^{*}$) the tunneling term is 
minimized, as yields $\theta =Qx$, or
%%%%%%%%%%%%%%%%%%%%%%%%%%%
\begin{equation}
\theta (x) = 2\pi xdB_{\parallel }/\phi _{\rm 0}  = 
2\pi xdB_{\perp }\tan\Theta /\phi _{\rm 0} .
\label{PhaseAngle}
\end{equation}
%%%%%%%%%%%%%%%%%%%%%%%%%%%
The phase difference $\theta (x)$ counts 
the number of flux penetrated into the area $xd$ of the junction.  
As the tilted angle increases, the screening currents 
$\left|J^{\text{f,b}}_{\rm x}\right|$
increase, 
and they will decrease the 
activation energy by destabilarizing excitations across the two layers.  
On the other hand, 
in the incommensurate phase ($\Theta >\Theta ^{*}$) the kinetic term is 
minimized, as yield $\theta =\theta _{\rm 0}=$ constant.  
No screening current flows, which means that the activation 
energy is insensitive to the tilted angle.  
The critical angle 
$\Theta ^{*}$ is 
given at the balanced point \cite{YangMoon} by 
$\tan\Theta ^{*}=({1/2\pi ^{2}d})\sqrt {\DSAS/\nt\rho _{\rm s}}$.  
It decreases as 
$\nt$ increases, as is qualitatively consistent 
with the data (Fig.\ref{Plateau1}).

\begin{figure}[!bth]
\epsfxsize=80mm
\epsfbox{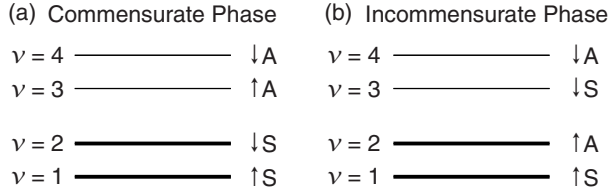}
\caption{\small
Schematic representation of the alignment of energy levels of electrons
at $\nu=2$.
The symmetric and antisymmetric states are represent by S and A.
The short vertical arrows represent 
the orientations of electron spin.
At $\nu=2$
charge excitations flip spins in the incommensurate phase.
On the contrary,
at $\nu=1$ they do not flip spins 
in the incommensurate phase.
}
\label{BLvacc0PS}
\end{figure}

An important observation is 
that $\cos(\theta _{\rm 0}-Qx)$ oscillates very rapidly 
in the incommensurate phase, and its average vanishes.  
Consequently, the tunneling energy is suppressed as a many-body effect, 
and the second energy level is given by the antisymmetric spin-up state 
at the balanced point [Fig.\ref{BLvacc0PS}(b)].  
Because charge excitations do not acquire the Zeeman energy, 
the activation energy is flat in the incommensurate phase at 
$\nu =1$ in the balanced configuration, 
as explain the data (for $\sigma =0$) in  
Fig.\ref{Activa1} and also the data by Murphy \etal \cite{Sheena}.

%%%%%%%%%%%%%%%%%%%%%%%%%%%%%
%%%%%%%%%%%%%%%%%%%%%%%%%%%%%

We turn to discuss the BLQH state at $\nu=2$.
When the total density is sufficiently small the interlayer coherence develops 
as in the $\nu=1$ state,  
and will be also described by the effective Hamiltonian (\ref{EffecHamil}).
Because the tunneling energy gap $\DSAS$ is 
larger than the Zeeman energy ($g^{*}\mu _{\rm B}B$) 
in our sample ($\DSAS/g^{*}\mu _{\rm B}B\simeq 4$ at $B=$ 5\,T), 
the lowest two levels occupied are the symmetric spin-up and 
spin-down states in the commensurate phase [Fig.\ref{BLvacc0PS}(a)].  
The activation energy decreases as 
the tilted angle increases as in the $\nu=1$ commensurate phase.  
In the incommensurate phase, since the tunneling 
interaction is effectively suppressed by a many-body effect, the lowest two 
levels are the symmetric and antisymmetric spin-up states 
[Fig.\ref{BLvacc0PS}(b)].  
Consequently, charge excitations flip spins, as is 
seen in the increase of the activation energy (for $\nt=0.7$ and $\sigma =0$) 
in Fig.\ref{Activa2}.

This physical interpretation of our magnetotransport experiments is 
consistent with the one obtained from the optical experiments 
\cite{PellegriniA,PellegriniB}, 
except that we have not observed their D phase.
We have argued that, when $\DSAS>g^{*}\mu _{\rm B}B$ at $\Theta =0$, 
the CIC phase transition rearranges the energy levels 
in the incommensurate phase (Fig.\ref{BLvacc0PS}).  
It induces effectively a transition 
between the spin polarized phase and 
the spin unpolarized phase at $\nu =2$.  
This is precisely the feature observed by the optical experiments 
\cite{PellegriniA,PellegriniB}.  
We summarize the relations between the results in these two 
types of experiments in TABLE \ref{TableA}.
Finally, we remark that the interlayer coherent state and the compound state
correspond to the canted state and the FM state 
in a theoretical work \cite{DasSarma} at $\nu=2$.

We thank T. Saku (NTT) for growing the sample used in the present work.
Part of this work was done at Laboratory for Electronic Intelligent Systems, 
RIEC, Tohoku University.
The research was supported in part by Grant-in-Aids for the Scientific 
Research from the Ministry of Education, Science, Sports and Culture 
(10203201, 09244103, 10138203).

%%%%%%%%%%%%%%%%%%%%%%%%%%%%%%%%%%%%%%%%%%%%%%%%%%%%%%%%%%%%%%%
\begin{table}[h]
\caption{
Comparison between the optical and magnetotransport results at $\nu =2$.}
\scriptsize
\begin{tabular}{l||l|l|l} \hline
       & Low Density  & Low Tilt\ \ \ \ \  & Low Temp. \\
Sample & $\dnA $          & $\dnA $       & $\downarrow$ \\
       & High Density & High Tilt          & High Temp. \\ \hline
       \hline
Our work       & Coherent    & Commens.          &  

\\
$\DSAS=6.8$\,K & $\dnA  \nt\simeq 0.9$ & $\dnA \Theta ^{*}\simeq  50^{\circ}$ & No detection 
\\
$\nt=0.6\sim 1.6$  & Compound    & Incommens.        &  

\\ \hline
Pellegrini \etal & Unpolarized & Unpolarized        & Unpolarized \\
$\DSAS\simeq 7$\,K     & $\dnA  \nt\simeq 1.3$ & $\dnA  \Theta ^{*}\simeq  37^{\circ}$ &$\dnA  T^{*}\simeq 0.5$\,K \\
$\nt=0.6\sim 1.4$    &Polarized    & Polarized          & D Phase \\ \hline
\end{tabular}
\label{TableA}
\end{table}
%%%%%%%%%%%%%%%%%%%%%%%%%%%%%%%%%%%%%%%%%%%%%%%%%%%%%%%%%%%%%%%

\scriptsize
%%%%%%%%%%%%%%%%%%%%%%%%%%%%%%%%%%%%%%%%%

\end{multicols}

\begin{thebibliography}{99}
%%%%%%%%%%%%%%%%%%%%%%%%%%
\bibitem{EIcoher}
Z.\ F.\ Ezawa and A.\ Iwazaki, Int.\ J.\ Mod.\ Phys.\  B {\bf 6}, 3205 (1992);
Phys.\ Rev.\ B \  {\bf 47}, 7295 (1993); 
$ibid$. {\bf 48}, 15189 (1993). 
%%%%%%%%%%%%%%%%%%%%%%%%%%
\bibitem{WenZee}
X.\ G.\ Wen and A.\ Zee,
Phys.\ Rev.\ Lett.\  {\bf 69}, 1811 (1992).
%%%%%%%%%%%%%%%%%%%%%%%%%%
\bibitem{Sheena}
S.Q. Murphy, J.P. Eisenstein, G.S. Boebinger, L.N. Pfeiffer 
and K.W. West, Phys.\ Rev.\ Lett.\  {\bf 72}, 728 (1994).
%%%%%%%%%%%%%%%%%%%%%%%%%%
\bibitem{EIplasmon}
Z.\ F.\ Ezawa, Phys.\ Rev.\ B \  {\bf 51}, 11152 (1995);
Z.\ F.\ Ezawa and A.\ Iwazaki, Int.\ J.\ Mod.\ Phys.\  B {\bf 8}, 2111 (1994).
%%%%%%%%%%%%%%%%%%%%%%%%%%
\bibitem{YangMoon}
K. Yang, K. Moon, L. Zheng, A.H. MacDonald, S.M. Girvin,
D. Yoshioka and S.C. Zhang,
Phys.\ Rev.\ Lett.\  {\bf 72}, (1994) 732;
K. Moon, H. Mori, K. Yang, S.M. Girvin, A.H. MacDonald,
L. Zheng, D. Yoshioka and S.C. Zhang,
Phys.\ Rev.\ B \  {\bf 51}, (1995) 5138.
%%%%%%%%%%%%%%%%%%%%%%%%%%
\bibitem{SawaPRLa}
A. Sawada, Z.F. Ezawa, H. Ohno, Y. Horikoshi, Y. Ohno,
S. Kishimoto, F. Matsukura, M. Yasumoto, and A. Urayama,
Phys.\ Rev.\ Lett.\  {\bf 80}, 4534 (1998).
%%%%%%%%%%%%%%%%%%%%%%%%%%
\bibitem{EzaICBa}
Z.\ F.\ Ezawa, Physica\ B {\bf 249--251}, 841 (1998);
Phys.\ Lett.\  A {\bf 249}, 223 (1998).
%%%%%%%%%%%%%%%%%%%%%%%%%%
\bibitem{PellegriniA}
V. Pellegrini, A. Pinczuk, B.S. Dennis, 
A.S. Plaut, L.N. Pfeiffer and K.W. West,
Phys.\ Rev.\ Lett.\  {\bf 78}, (1997) 310.
%%%%%%%%%%%%%%%%%%%%%%%%%%%%%%%%%%%%%%%%%%%%%%%%%%%%%%%%%%%  
\bibitem{PellegriniB}
V. Pellegrini, A. Pinczuk, B.S. Dennis, 
A.S. Plaut, L.N. Pfeiffer and K.W. West,
Science\  {\bf 281}, (1998) 799.
%%%%%%%%%%%%%%%%%%%%%%%%%%%%%%%%%%%%%%%%%%%%%%%%%%%%%%%%%%%  
\bibitem{GONO}
M. Suzuki, A. Sawada, A. Ishiguro and K. Maruya,
Cryogenics {\bf 37}, (1997) 275.
%%%%%%%%%%%%%%%%%%%%%%%%%%
\bibitem{SSC}
A. Sawada, Z.F. Ezawa, H. Ohno, Y. Horikoshi, O. Sugie,
S. Kishimoto, F. Matsukura, Y. Ohno and M. Yasumoto,
Solid\ State\ Commun.\  {\bf 103}, 447 (1997).
%%%%%%%%%%%%%%%%%%%%%%%%%%
\bibitem{EzaIQC}
Z.\ F.\ Ezawa, Phys.\ Lett.\  A {\bf 229}, 392 (1997);
Phys.\ Rev.\ B \  {\bf 55}, 7771 (1997).
%%%%%%%%%%%%%%%%%%%%%%%%%%
\bibitem{DasSarma}
L. Zheng, R.J. Radtke and S. Das Sarma, Phys.\ Rev.\ Lett.\  {\bf 78}, 
(1997) 2453.
%%%%%%%%%%%%%%%%%%%%%%%%%%
\end{thebibliography}
\end{document}